%% file: main.tex
\documentclass[conference]{IEEEtran}
\IEEEoverridecommandlockouts
\usepackage{array}
\usepackage{multirow}
\usepackage{cite}
\usepackage{amsmath,amssymb,amsfonts}
\usepackage{algorithmic}
\usepackage{graphicx}
\usepackage{textcomp}
\usepackage{balance}
\usepackage{xcolor}
\usepackage{cases}

\makeatletter
\def\ps@IEEEtitlepagestyle{%
  \def\@oddfoot{\mycopyrightnotice}%
  \def\@evenfoot{}%
}
\def\mycopyrightnotice{%
  {\footnotesize 978-1-6654-4331-9/21/\$31.00 \copyright 2021 IEEE \hfill}
  \gdef\mycopyrightnotice{}
}

\usepackage{subfigure}
\def\BibTeX{{\rm B\kern-.05em{\sc i\kern-.025em b}\kern-.08em
    T\kern-.1667em\lower.7ex\hbox{E}\kern-.125emX}}
\usepackage{tikz}    
    
\begin{document}

\title{Q-learning-based Opportunistic Communication for Real-time Mobile Air Quality Monitoring Systems}

\author{
    \IEEEauthorblockN{Trung Thanh Nguyen\IEEEauthorrefmark{1}, Truong Thao Nguyen\IEEEauthorrefmark{2}, Tuan Anh Nguyen Dinh\IEEEauthorrefmark{1}, Thanh-Hung Nguyen\IEEEauthorrefmark{1}, Phi Le Nguyen\IEEEauthorrefmark{1}}
    \IEEEauthorblockA{\IEEEauthorrefmark{1}School of Information and Communication Technology, Hanoi University of Science and Technology, Hanoi, Vietnam
    \IEEEauthorblockA{\IEEEauthorrefmark{2}The National Institute of Advanced Industrial Science and Technology (AIST), Japan}
    \{thanh.nt176874@sis, anh.ndt164767@sis, hungnt@soict, lenp@soict\}.hust.edu.vn; nguyen.truong@aist.go.jp}
}

\newcommand{\nguyen}[1] { {\color{blue}{#1}}}
\newcommand{\thanhhff}[1] { {\color{red}{#1}}}
\maketitle
\begin{abstract}
We focus on real-time air quality monitoring systems that rely on devices installed on automobiles in this research.
We investigate an opportunistic communication model in which devices can send the measured data directly to the air quality server through a 4G communication channel or via Wi-Fi to adjacent devices or the so-called Road Side Units deployed along the road.
We aim to reduce 4G costs while assuring data latency, where the data latency is defined as the amount of time it takes for data to reach the server.
We propose an offloading scheme that leverages Q-learning to accomplish the purpose. The experiment results show that our offloading method significantly cuts down around 40-50\% of the 4G communication cost while keeping the latency of 99.5\% packets smaller than the required threshold.
\end{abstract}


\section{Introduction}
\input{1-introduction} \label{sec:introduction}


\section{Preliminaries}
\input{2-preliminaries}


\section{Proposal} \label{sec:fuzzy_q_learning}
\input{4-fuzzyQlearning}

\section{Evaluation} \label{sec:experimental}
\input{5-experimentalResults}
\section{Conclusion} \label{sec:conclusion_and_future_work}
\input{6-conclusion}

\section*{Acknowledgment}
This work was supported by JST, ACT-X Grant Number JPMJAX190C, Japan.
This research is funded by Vietnam National Foundation for Science and Technology Development (NAFOSTED) under grant number 102.01-2020.23.

\balance
\addtolength{\textheight}{-12cm}   
\bibliographystyle{unsrt}
\bibliography{ref}
\balance

\end{document}

%% file: 1-introduction.tex
In the last decades, with the rapid development of industrialization and urbanization, air pollution has become an increasingly crucial issue.
Over the years, many studies have been conducted and showed that air pollution could cause diseases, allergies, and even death to humans \cite{CHEN2019740, LI20191304}.
In that context, monitoring air quality is one of the critical factors that help the government make policies and people plan for life.
Traditionally, air quality monitoring has been carried out by static monitoring stations located at fixed locations.
However, due to the high deployment and operating costs, the density of deployed monitoring stations is insufficient. 
For example, in Hanoi, Vietnam, with more than 3000 $km^2$, there are only $50$ air quality monitoring stations \cite{PAM_Air}.

Recently, there have been several studies proposing a new approach, namely the vehicle-based mobile air quality monitoring system \cite{KAIVONEN202023} \cite{9322079}.
The vehicle-based mobile air quality monitoring systems leverage lightweight air quality monitoring devices mounted on vehicles to broaden the monitoring area. 
In \cite{DESOUZA2020102239}, the authors considered a mobile air quality monitoring system that relies on low-cost mobile sensors deployed on trash trucks. They proposed techniques to detect pollution hot spots and identify pollutant source signatures.
Nguyen et al. in \cite{9322079} studied how to deploy air quality monitoring sensors on buses for maximizing the monitored regions. The authors first mathematically formulated the problem and then provided an approximation algorithm to determine optimal buses for placing sensors.

In this research, we focus on real-time vehicle-based mobile air quality monitoring systems, in which the devices continuously collect the air quality information and transfer it to the server. 
There are two challenges when dealing with such real-time monitoring systems.
Firstly, we need to ensure the freshness of the information, i.e., guaranteeing that the time from when the data is measured till arriving at the server does not exceed a threshold. 
The second challenge is to minimize the communication cost. 
The targeted problem, which we name as OCMA (stands for Opportunistic Communication for Mobile Air quality monitoring) can be stated as follows. 
Air quality monitoring devices are mounted on buses. These devices perform air quality measurement with frequency $f$. The data collected by the devices are transmitted to the cloud server via one of the following communication planes.
Firstly, devices can transmit data directly to the cloud server via the 4G communication.
Secondly, the devices can transmit data to Road Side Units (RSUs) located along the roads through the Wi-Fi channel. These Road Side Units will transfer data to the cloud server through a high speed wired network.
Finally, a device can relay data to another device on the vehicle next to it. This neighbor vehicle will then aggregate the data and transfer it to the cloud server or Road Side Units.
We assume that the 4G communication is available everywhere. Thus, device can transmit data by 4G at any time. In contrast, Wi-Fi communication can only be used when devices enter the communication ranges of the RSUs or other devices. 
On the other hand, it is well-known that 4G communication is usually much more expensive than Wi-Fi.
Besides, the 4G communication also consumes much more energy compared to Wi-Fi. 
Therefore, our OCMA problem asks to minimize the use of 4G communication while guaranteeing that the information latency does not exceed a threshold. Here, the term ``information latency" is defined by the time interval from when the data is collected until it reaches the server. 
To the best of our knowledge, this study is an early attempt to minimize communication costs while maintaining the freshness of information in mobile air quality monitoring systems.

The OCMA problem can be categorized as an offloading problem in V2X (i.e., Vehicle-to-Everything) networks. 
In \cite{7553459}, K. Zhang et al. addressed the energy optimization problem in Mobile Edge Computing (MEC)-enabled 5G networks. They first mathematically formulated the targeted problem and then proposed an approximation algorithm to allocate the radio resource.  
The work in \cite{platoon_1, platoon_3, platoon_2} addressed the offloading decision of collaborative task execution between platoons and a MEC server. Both \cite{platoon_1, platoon_3} considered how to determine the location of task execution either on a vehicle, offloading to  other platoon members, or an associated MEC server. However, \cite{platoon_1} focused on minimizing the offloading cost, while \cite{platoon_3} aimed at reducing the average energy consumption. \cite{platoon_2} targeted the reduction of offloading latency between the vehicles, each of which necessarily maintains the information of its 1-hop neighbors. Whenever with a task, a vehicle calculates the offloading latency for all the relay hop candidates. A neighbor vehicle with minimum latency is chosen as the best relay node. 
The authors in \cite{8644315} proposed a federated offloading method that exploits  horizontal offloading path between  vehicles, with the objective of minimizing total latency.
In \cite{2_tier_1}, the authors aimed at minimizing the power consumption of MEC servers and vehicles.

Zhao et al. recently utilized MEC and cloud computing resources simultaneously for offloading \cite{3-tier_1}. In that work, vehicles could offload their computation tasks to an MEC server or the cloud via RSUs. The objective was to maximize the system's utility by optimizing both the offloading strategy and resource allocation. In \cite{8402110}, Y. Lin et al. addressed the traffic and capacity allocation problem in a three-tier model and proposed an optimization algorithm consisting of two phases. The first was adjusting the capacity allocation, optimizing the traffic allocation. Their objective was to minimize total capacity and guarantee that at least some traffic has satisfying latency constraints. 

Unlike previous research, we use three communication planes simultaneously, namely, vehicle-to-cloud, vehicle-to-RSU, and vehicle-to-vehicle, with the goal of reducing 4G communication costs while maintaining information freshness. Our idea is to exploit Q-learning in offloading tasks. 
In our Q-learning paradigm, each air quality monitoring device is considered as an agent that keeps track of its Q-table.
The action space consists of four values: keeping data in the local memory, sending to the RSU, relaying to the neighbor device, and transmitting to the server.  
An entry in a Q-table represents the quality of an action. 
The agent (i.e., air quality monitoring device) chooses the action whose Q-value is the greatest at every time slot. 
The Q-table is updated after performing an action, based on a so-called reward function. Our reward function is designed to encourage actions that reduce 4G communication cost while ensuring the data latency constraint. 
Our contribution is as follows:
\begin{itemize}
    \item We are the first to handle the challenge of opportunistic communication in real-time mobile air quality monitoring systems.  
    \item We propose a Q-learning-based opportunistic communication protocol that attempts to reduce the cost of 4G communication while ensuring data latency. 
    \item We perform extensive experiments to evaluate the performance of the proposed protocol. 
    The results show the superiority of our proposal to the existing works. 
\end{itemize}

The remainder of the paper is organized as follows. 
We present the network model and a brief introduction to Q-learning framework in Section \ref{sec:preliminaries}.
Sections \ref{sec:fuzzy_q_learning} describes our proposed protocol in details. 
We present the numerical results in Section \ref{sec:experimental} and conclude the paper in Section \ref{sec:conclusion_and_future_work}.

%% file: 2-preliminaries.tex
\label{sec:preliminaries}
In this section, we first present our network model, and then we give a brief introduction to Q-learning, the technique that will be used in our solution.

\subsection{Network Model} 
\begin{figure}
    \centering
    \includegraphics[width=0.8\columnwidth]{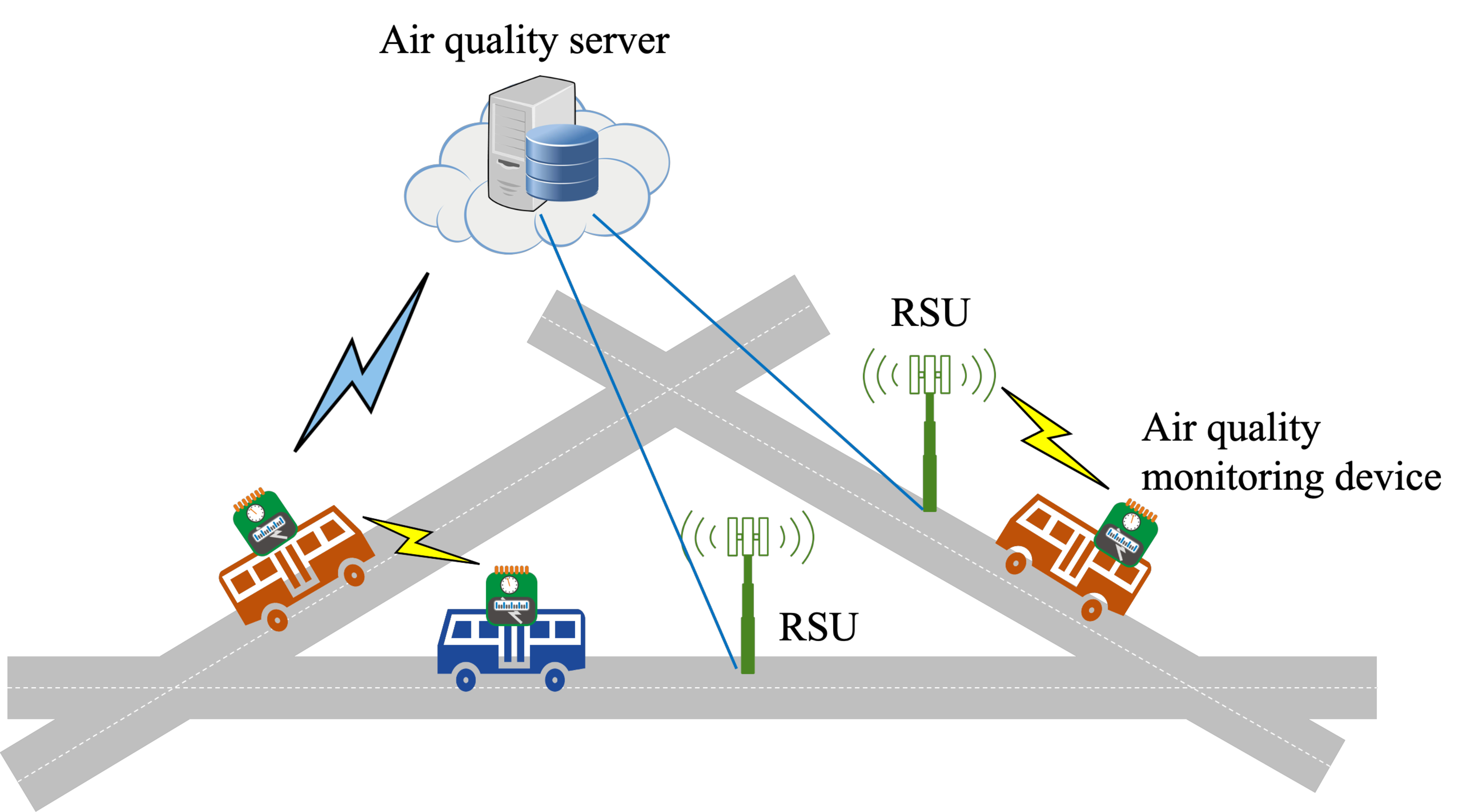}
    \caption{Network model.}
    \label{fig:network_model}
\end{figure}
Figure \ref{fig:network_model} depicts the network model which comprises of three components: air quality monitoring devices, Road-Side-Units (RSUs), and a cloud server. 
We assume that there are $n$ air quality monitoring devices that are mounted on $n$ buses. 
Each device $D_i (i = 1,\dots,n)$ has a computing capacity of $C^{*}_{i}$ and the transmission range of $r_{D_i}$. 
The RSUs are computing units located along the roads.
We also assume that there are $m$ RSUs denoted as $R_j (j = 1,\dots,m)$ which has the transmission range of $r_{R_j}$.
The cloud server is named as the \emph{air quality server}.
The air quality monitoring devices continuously measure the air quality indicators and transfer the measured data to the air quality server via one of the following three communication planes:
\begin{enumerate}
\item Air quality monitoring device $\rightarrow$ Air quality server.
\item Air quality monitoring device $\rightarrow$ RSU $\rightarrow$ Air quality server.
\item Air quality monitoring device A $\rightarrow$ Air quality monitoring device B (device B then transfers the data to a RSU or the air quality server, or another device). 
\end{enumerate}

The air quality monitoring devices use the 4G connection, which is relatively expensive, to interact with the air quality server.
Communication between air quality monitoring devices and RSUs, on the other hand, takes place via a free Wi-Fi channel, as does communication between two air quality monitoring devices.
It's worth noting that 4G service is available everywhere. As a result, a device can send data to a server via the 4G channel nearly instantly. In order to send data to the RSU or another device, the device must first move into that RSU's or device's communication range. 
We aim at proposing an offloading mechanism that accomplishes both of the following goals: 1) Reducing the amount of data with a latency larger than a given threshold $\delta$, where ``data latency" refers to the time it takes for data to reach the server from when it is measured; and 2) Minimizing the amount of data transmitted by 4G communication.

\subsection{Q-learning}

\begin{figure}
    \centering
    \includegraphics[width=5cm]{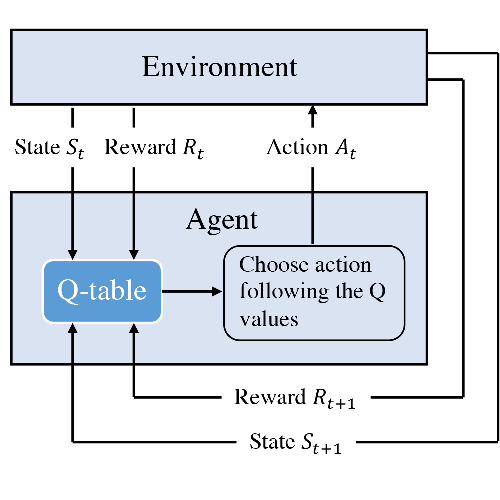}
    \caption{Q-learning overview.}
    \label{fig:q_learning}
    \vspace{-0.3cm}
\end{figure}

Q-Learning \cite{watkins1992q} is a Reinforcement Learning technique that learns from experiences in order to achieve a certain optimization target. Figure \ref{fig:q_learning} depicts the Q-learning process, which consists of four components: an environment, an agent, a state space, and an action space.
The agent uses the so-called Q table to determine the action at each stage. 
Each entry in the Q table represent the goodness of performing an action at a given state concerning the agent's final goal. 
The Q-table is updated by the following Bellman equation: 
\begin{equation}
\small
\label{eq}
Q(S_t, A_t) \leftarrow (1 - \alpha)Q(S_t, A_t) + \alpha[\mathcal{R}_t + \gamma\max_{a}Q(S_{t+1}, a)]
\end{equation}
where $Q(S_t, A_t)$ is the Q-value at a state $S_t$ when taking action $A_t$ at time-step $t$. $\mathcal{R}_t$ is the reward received for performing action $A_t$ in the state of $S_t$. $\max_{a}Q(S_{t+1}, a)$ is the maximum value that may be obtained for all possible actions $a$ at the next state $S_{t+1}$. In addition, the $\alpha$ and $\gamma$ (ranges from $0$ to $1$) represent the learning and discount rates, respectively. 

%% file: 4-fuzzyQlearning.tex
In our Q-learning-based model, the network is considered the environment, while each air quality monitoring device is an agent.
We utilize the distributed approach where each monitoring device runs its own Q-learning-based model. 
To facilitate the reading, we summarize the notations in Table \ref{tab:notions}. 
\begin{table}[tb]
    \centering
    \caption{Notions \label{tab:notions}}
    \begin{tabular}{|m{1.2cm}|m{6.7cm}|}
    \hline
         \emph{Notion} & \emph{Description} \\
         \hline
         $n$ & the number of air quality monitoring devices \\
         $m$ & the number of RSUs \\
         $D_i$ & the $i$-th air quality monitoring device \\
         $r_{D_i}$ & the transmission range of $D_i$  \\
         $C^{*}_i$ & the maximum computing capacity of $D_i$\\
         $R_j$ & the $j$-th RSU \\ 
         $r_{R_j}$ & the transmission range of $R_j$ \\
         $\delta$ & the data latency threshold\\
         $S_i(t)$ & the state at a time slot $t$ of agent $D_i$\\
         $A_i(t)$ & the action taken by agent $D_i$ at a time slot $t$\\
         $\mu_i(t)$ & the timing agent $D_i$ generates the last data at a time slot $t$\\
         $c_i(t)$ & the remaining computing resource of $D_i$ at time slot $t$ \\
         $c_\mathcal{N}(t)$ & the remaining computing resource of the nearest device at time slot $t$ \\
         $\Delta_i(t)$ &  the time interval from when $D_i$ generates the last data until the time slot $t$,  $\Delta_i(t)= t - \mu_i(t)$\\
         $\Delta{c}(t)$ & the difference in remaining capacity of $D_i$ and nearest device at time slot $t$, $\Delta{c}(t) = c_{\mathcal{N}}(t) - c_i(t)$\\
         $\theta$ & the priority factor\\
         \hline
    \end{tabular}
    \vspace{-0.2cm}
\end{table}
\subsection{State space}
For each device $D_i$, the state at a time slot $t$ is a quadruple consisting of the following items: 
\begin{itemize}
  \item $\mu_i(t)$: the timing $D_i$ generates the last packet.
  \item $c_i(t)$: the computing resource of $D_i$ that is remaining at time slot $t$. 
  \item $c_\mathcal{N}(t)$: the remaining resource of the nearest device $\mathcal{N}$, if $\mathcal{N}$ is in the communication range of $D_i$.
  \item $\mathcal{N}_i^R(t)$: a binary variable indicating whether $D_i$ is in the communication range of a RSU.
\end{itemize}

\subsection{Action space} 
An air quality monitoring device can conduct one of the following actions at each time slot $t$:
\begin{enumerate}
    \item[i)] Keeping the data in the local queue,
    \item[ii)] Sending the data directly to the air quality server via 4G communication channel, 
    \item[iii)] Sending the data to the nearest RSU, if $D_i$ is in the communication range of a RSU,
    \item[iv)] Sending the data to the nearest device, if they are in the communication range of each other. 
\end{enumerate}

\subsection{Reward function} 
We denote by $\mathcal{R}_i(t)$ the reward received when device $D_i$ performs action $A_i(t)$. 
Our goal is to minimize the total amount of data transmitted by 4G while guaranteeing that the data latency does not exceed a predefined threshold $\delta$.
The reward function is defined as follows.

\begin{numcases}{
\small
\mathcal{R}_i(t)=}
    \frac{\theta \times C^*_i - c_i(t)}{1 + \Delta_i(t)} & \text{, sending to the} \nonumber \\
    & \text{air quality server} ~\label{eq1}\\ 
\frac{C^*_i - \theta \times c_i(t)}{1 + \Delta_i(t)} & \text{, sending to a} \nonumber\\ 
& \text{RSU} ~\label{eq2}\\
\frac{\Delta{c}(t)}{[1 + \Delta_i(t)] \times \left | \Delta{c}(t) \right |} & \text{, sending to the} \nonumber\\
& \text{nearest device} ~\label{eq3_4} \\ 
-p & \text{, if} \ $c_i(t) > C^*_i$  \nonumber\\
& \text{or} \ $\Delta_i(t) > \delta$ ~\label{eq5}\\
0 & \text{, keeping in the} \nonumber\\
& \text{local memory} ~\label{eq6}
\end{numcases}
where $\Delta_i(t) = t - \mu_i(t)$ is the time elapsed from when the data is collected, $\Delta{c}(t) = c_{\mathcal{N}}(t) - c_i(t)$ in that $c_{\mathcal{N}}(t)$ is the remaining resource of the nearest device, and $p$ is a significantly large positive number.
$\theta$ is a parameter in the range of $[0, 1]$. 
The value of $\theta$ is adjust based on the remaining capacity $c_i(t)$ and the elapsed time $\Delta_i(t)$.
Specifically, when $c_i(t)$ is large and $\Delta_i(t)$ is small, $\theta$ tends to be small as sending the packet to RSU is prioritized than sending it to the gNB. 
The rationale behind the reward function is as follows. 
Formula (\ref{eq1}) means that when the device's remaining capacity is sufficient ($\theta \times C^*_i - c_i(t) \leq 0$), the device will avoid transmitting packets directly to the cloud server (due to the negative reward value). Instead, it will either send the data to the RSU or a nearest device, or hold the packet in local memory until it can take another action.
In contrast, when the remaining capacity is no longer sufficient ($\theta \times C^*_i - c_i(t) > 0$), the device cannot hold the packet in the local memory; thus, transmitting the data directly to the cloud server gets more priority. 
Moreover, when $\theta \times C^*_i - c_i(t) \leq 0$, the reward of action sending to the cloud server is proportional to $\Delta_i(t)$.
It means that the smaller the $\Delta_i(t)$, the more likely the device will not send the packet to the air quality server. 
The reward of action transferring data to a RSU is represented by Formula (\ref{eq2}), which is inversely proportional to $\Delta_i(t)$.
When $\Delta_i(t)$ is small, the device will prioritize delivering data to the RSU.
When $\Delta_i(t)$ is significant large, however, communicating over the RSU is no longer a viable alternative because the device is not always within the RSU's communication range. Therefore, the reward of this action is lowered. 
Formulas (\ref{eq3_4}) mean that the action of relaying data to the neighboring device is only encouraged when the available resource of the neighboring device is greater than that of the current device. 
Moreover, the greater the $\Delta_i(t)$, the smaller the reward of action sending to the neighbor device (because $\frac{1}{1 + \Delta_i(t)}$ is inversely proportional to $\Delta_i(t)$). 
Finally, Formula (\ref{eq5}) depicts that when the available resource of the current device exceeds its capacity or when the data's latency exceeds the threshold, the agent will be punished by a substantial negative reward.

%% file: 5-experimentalResults.tex
\subsection{Methodology}
\label{sec:methodology}
In this section, we evaluate the efficiency of our proposed algorithm in terms of optimizing the communication performance and cost.

\textbf{Simulation model:} we simulate the target network model (as shown in Figure~\ref{fig:network_model}) by extending the queue-model proposed in \cite{Khiem_Le_MEC}. In which, air quality monitoring devices (sensors) iteratively generate packets with a same size in every $\lambda_d$ time steps. Packets are then stored in a local queue and wait for processing in a first-in-first-out manner. If the queue is full at the time a packet is generated, the packet will be dropped.
Thus, the \textit{data latency} of a packet comprises two components: the transmission latency, e.g., the total time for transmitting packets from the vehicles to the server, and the time in the queue of this packet. The transmission latency on a given communication channel, i.e., Wi-Fi, 4G, or wired network, is proportional to the packet size and inversely proportional to the link bandwidth. 
In our simulation,
the packet will be routed based on the corresponding decision algorithms, e.g., our proposed Q-learning method or the baseline method that uses random selection.
Basically, a packet is decided between four actions: (i) keep in the local queue, (ii) send directly to the air quality server, (iii) send to the nearest RSU, or (iv) send to the nearest sensor/vehicle. In case a packet of a given sensor is decided to send to the nearest RSU (or sensor) while there is no available RSU (or sensor) in the transmission range of this sensor, the packet will continue keeping in the local queue for processing in the next time step (hereafter mentioned as the \textit{offload-hit}) or sent directly to the server when the queue remaining capacity is not sufficient (hereafter mentioned as the \textit{offload-missed} issue).

\begin{figure}
    \centering
    \includegraphics[width=0.35\textwidth]{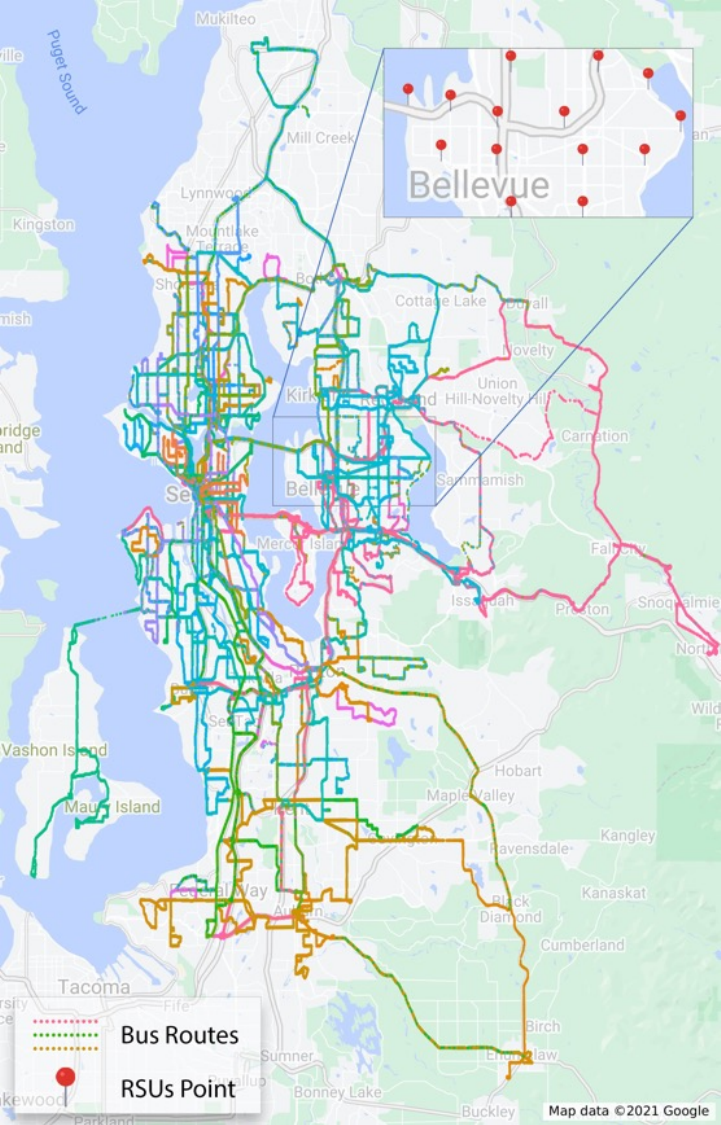}
    \caption{The real vehicle trajectory.}
    \label{fig:city_map}
    \vspace{-0.2cm}
\end{figure}

\textbf{Simulation environment:} we use the data set of bus routes in  Seattle City, Washington \cite{jetcheva2003design} to simulate the movement of vehicles. Each data point includes the time and position of a bus. 
We use the data collected within 2 days (48 hours) from November 19th, 2001 till the end of November 20, 2001. We realized that in this data set, there are several buses that only appear in a short time, therefore we only collect data of buses whose active time is no less than 90 minutes per day.  
We then generate the RSUs' position on the map along each bus route. In which, the RSUs are concentrated in the city center, 1 - 3 km apart, while in the suburbs there will be a sparser number of RSUs, 4 - 8 km apart. We illustrate the bus routes (color lines) and the RSUs' positions (red pin) in Figure \ref{fig:city_map}. 

\textbf{Evaluation metrics:} It is worthy to note that the target of this work is to keep the number of packets that can reach the air quality server as much as possible (i.e., \textit{maximizing the delivery ratio}), while reducing the amount of data with a long latency (i.e., \textit{guaranteeing the information freshness}) and lessening the 4G communication usage rates (i.e., \textit{ minimizing the communication cost}). 
Firstly, for the delivery ratio, we introduce the term rate of dropped packets ($r_{drop}$), i.e., the relative number of packets that have been dropped at sensors when their remaining capacity is no longer sufficient.
Secondly, for estimating the information freshness, we introduce the term $\delta$-delayed packets. The $\delta$-delayed packets are calculated by the total number of packets that have data latency greater than the threshold $\delta$. We also define the evaluation metric \textit{rate of $\delta$-delayed packets} (or $r_{delay}$) as the ratio of $\delta$-delayed packets over the total number of generated packets. 
Finally, the communication cost can be considered as the number of packets that directly send from a sensor to the air quality server via the 4G communication channel over the total number of generated packets (\textit{4G communication ratio} or $r_{server}$)\footnote{Assumption: the cost of sending a packet via the 4G communication channel is much higher than that of using Wi-Fi or wired network.}. 
In this work, we also investigate the ratio of a packet sent from a RSU to the air quality server, denoted as $r_{rsu}$.

\begin{table}[tb]
     \centering
     \caption{Configuration of the fix-possibility strategies}
     \label{tab:prob}
     \begin{tabular}{|c|c|c|c|c|c|}
         \hline
         Strategies & $P_{keep}$ & $P_{server}$ & $P_{rsu}$ & $P_{sensor}$\\
         \hline
         FP1 &0.2 & 0.3 & 0.3 & 0.2\\
         \hline
         FP2 &0.1 & 0.3 & 0.5 & 0.1\\ 
         \hline
         FP3 &0.1 & 0.5 & 0.3 & 0.1\\
         \hline
     \end{tabular}
\end{table}

\begin{table}[tb]
    \centering
    \caption{Simulation parameters \label{tab:params}}
    \begin{tabular}{| l | c |}
    \hline
         Parameter & Value \\
         \hline
         Packet size & 1 Mb\\
         RSU transmission range & 350 Meter \\ 
         Sensor transmission range & 120 Meter \\
         RSU-server's link bandwidth (wired network) & 10 Gbps \\
         Sensor-RSU's link bandwidth (wifi network) & 1 Gbps \\
         Sensor-server's link bandwidth (4G communication)  & 500 Mbps \\
         A time step $\mathcal{T}$ & 1 Min \\ 
         Packet generation interval at a sensor ($\lambda_d$) & 1 $\sim$ 5 $\mathcal{T}$ \\
         Data latency threshold $\delta$ & 5 $\sim$ 25 $\mathcal{T}$ \\
         Sensor's computing capacity $C^*$ & 25 Mb \\ 
         Number of RSUs $m$ & 384 \\ 
         Number of vehicles $n$ & 776 \\ 
         \hline
    \end{tabular}
\end{table}

\textbf{Comparison baseline:} Because there is no current work that handles the same problem as ours, to show the efficiency of our proposed method, we compare it with a naïve offloading strategy named \textbf{FP}. In FP, at a given time step, a packet is randomly decided between four actions, namely keeping at the local, sending directly to the server, transferring to an RSU, and relaying to the nearest device, with fixed possibilities of $P_{keep}$, $P_{server}$, $P_{rsu}$ and $P_{sensor}$, respectively. 
We consider three different configurations of FP, in which we change values of $P_{server}$, $P_{rsu}$ and $P_{sensor}$. 
The detail settings of FP are is summarized in Table~\ref{tab:prob}.

In the following, we first compare the performance and cost of our proposed method with the baseline concerning a particular settings of the packet generation interval $\lambda_d$ and the latency threshold $\delta$ in Section \ref{sec:compare}.
We then investigate the impacts of $\lambda_d$ and $\delta$ in Section \ref{sec:discussion}. 

\subsection{Comparison of the proposed method and the baseline}
\label{sec:compare}
In this experiment, we set the packet generation interval $\lambda_d$ to $1$, and the data latency threshold $\delta$ to $5$ and $10$. 
As mentioned in Section~\ref{sec:fuzzy_q_learning},
the value of $\theta$ is adjust based on the remaining capacity $c_i(t)$ and the elapsed time $\Delta_i(t)$. We heuristically increase $\theta$ from 0 to 1 when the remaining capacity decreases and the elapsed time increases.
Other simulation parameters are summarized in  Table \ref{tab:params}. 

\begin{figure}
    \centering
    \includegraphics[width=0.48\textwidth]{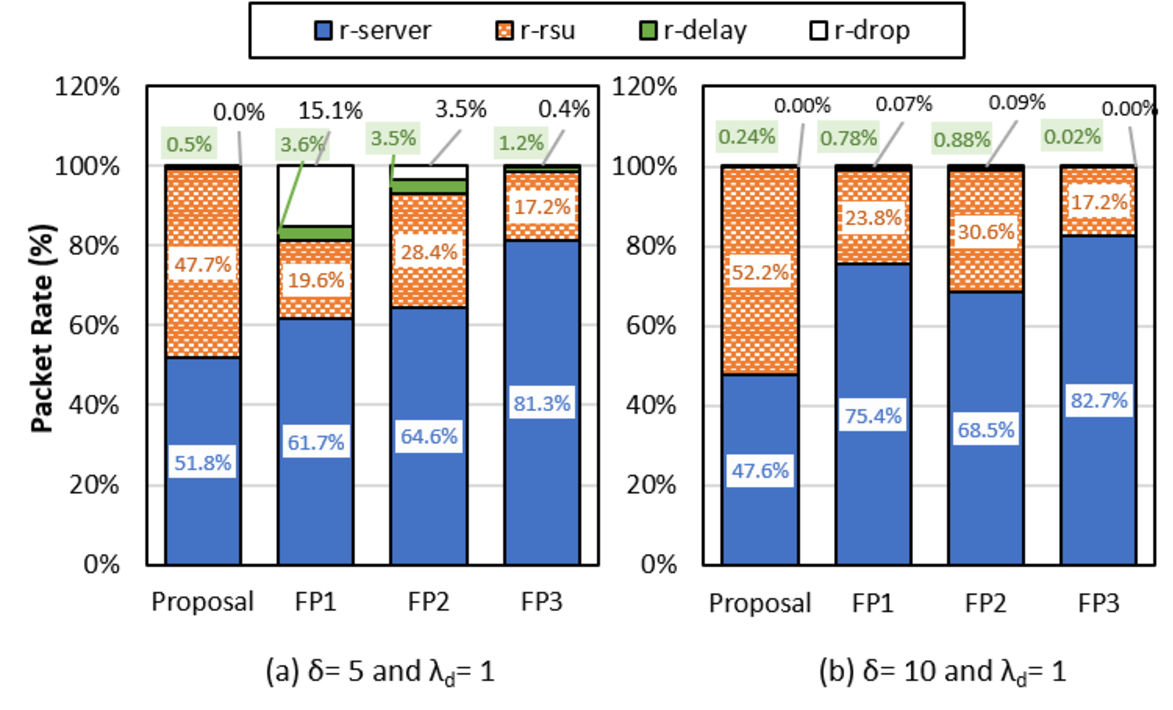}
    \caption{Relative breakdown of the simulation packets.}
    \label{fig:breakdown} 
    \vspace{-0.2cm}
\end{figure}

Figure~\ref{fig:breakdown} shows the rate of dropped packets $r_{drop}$, $\delta$-delayed packets $r_{delay}$, and the communication cost $r_{server}$ of our proposed method and the baseline strategies. The lower value the better performance.
First of all, in the baseline, devices tend to hold the data in their queue until they can send it an RSU/or the nearest device (the \textit{offload-hit} as mentioned in section~\ref{sec:methodology}). This strategy leads to a longer delay of a message and a higher number of dropped packets (when the queue is full). In contrast, by constructing a flexible priority factor $\theta$ which considers both the number of packets in the
local queue and the elapsed time of packets, our proposed method can always guaranty all the generated packets can reach the server, i.e., $r_{drop} = 0$. As a result, the baseline strategies can not avoid the packet-dropped issue in all the cases, e.g., 15,1\% and 3.5\% of $r_{drop}$ as in FP1 and FP2, respectively.
The result also shows that our method provides a small number of $\delta$-delayed packet, e.g., 0.5\% and 0.24\% which are $3-5\times$ lower than those of FP1 and FP2 as shown in Figure. \ref{fig:breakdown}(a) and \ref{fig:breakdown}(b), respectively. 

\begin{figure*}[tb]
    \centering
   \includegraphics[width=0.35\textwidth]{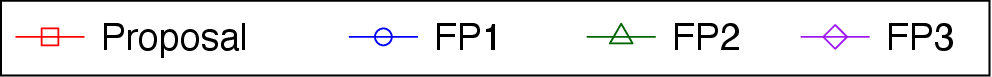}\\
    \subfigure[4G communication ratio]{\includegraphics[width=0.32\textwidth]{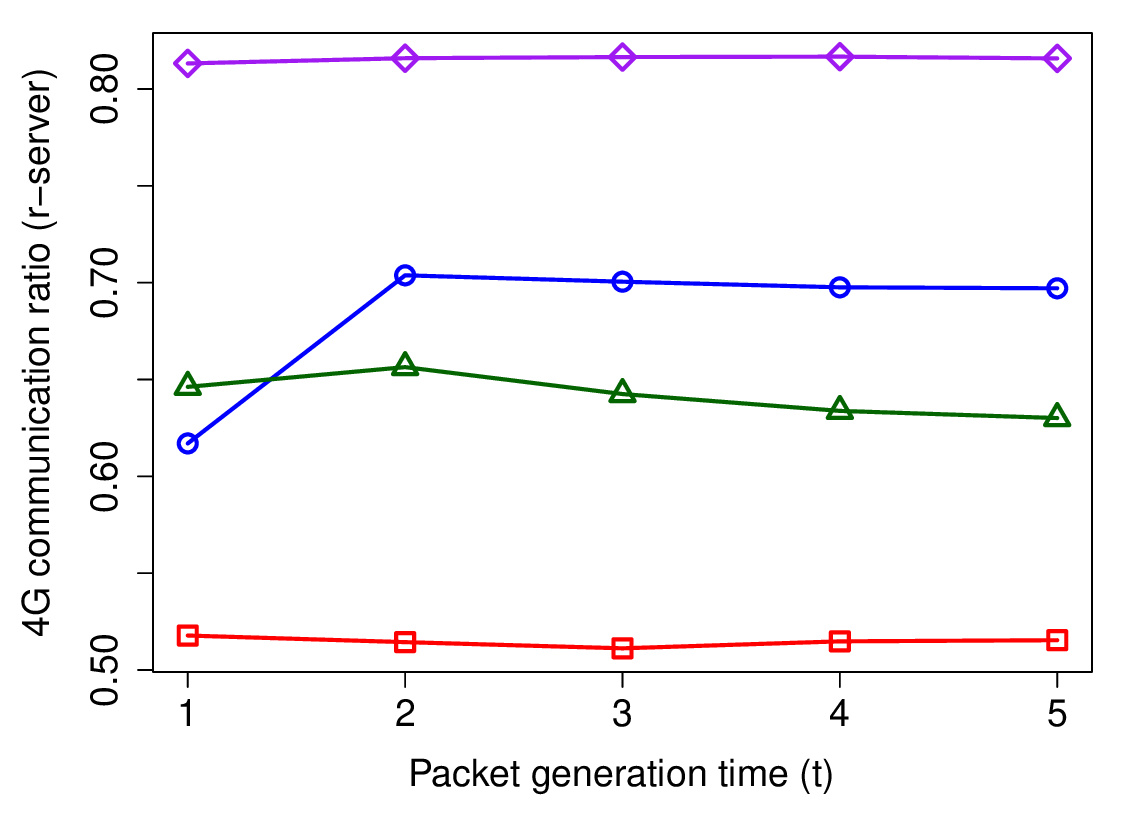}\label{fig:generation_time_1}}
    \hfill
    \subfigure[Rate of $\delta$-delayed packets]{\includegraphics[width=0.32\textwidth]{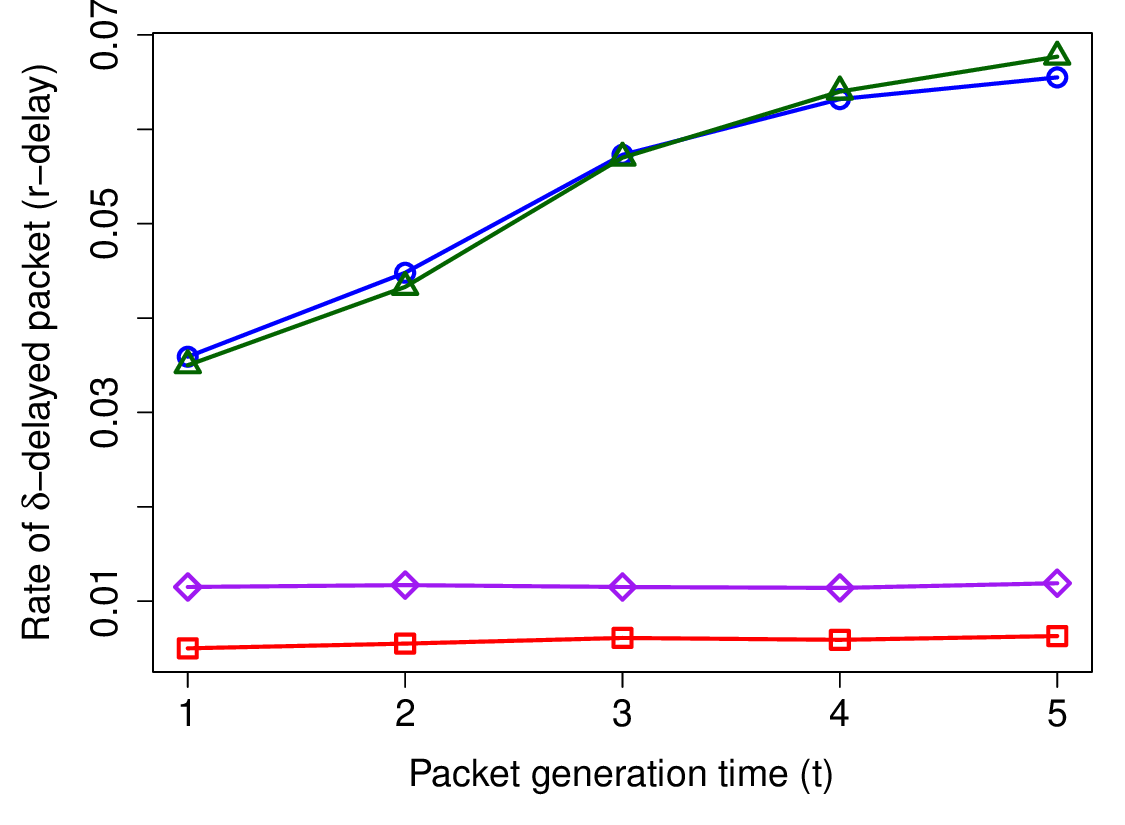}\label{fig:generation_time_2}}
     \hfill
    \subfigure[Rate of dropped packets]{\includegraphics[width=0.32\textwidth]{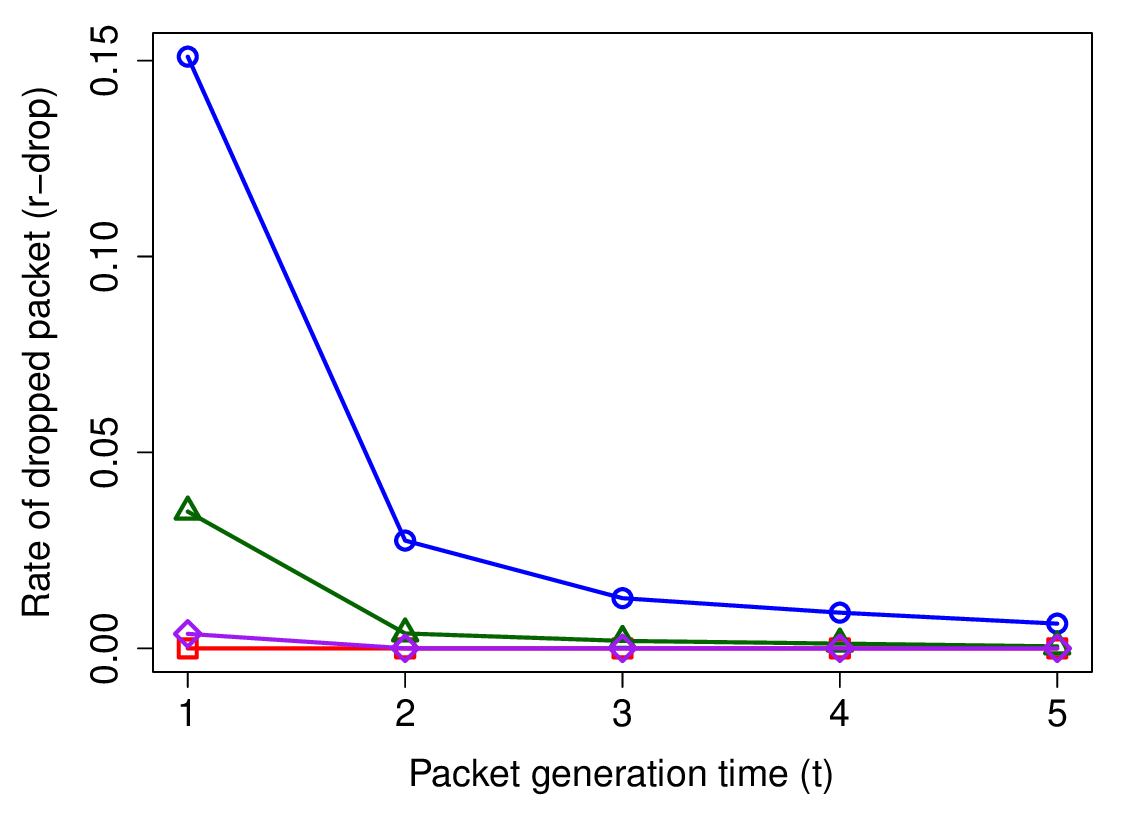}\label{fig:generation_time_3}}
    \caption{Impacts of the packet generation interval ($\lambda_d$). $\delta$ is fixed to 5 time steps. \label{fig:generation_time}} 
\end{figure*}
\begin{figure*}[tb]
    \centering
    \includegraphics[width=0.35\textwidth]{figs/r/legend_final.eps}\\
    \subfigure[4G communication ratio]{\includegraphics[width=0.31\textwidth]{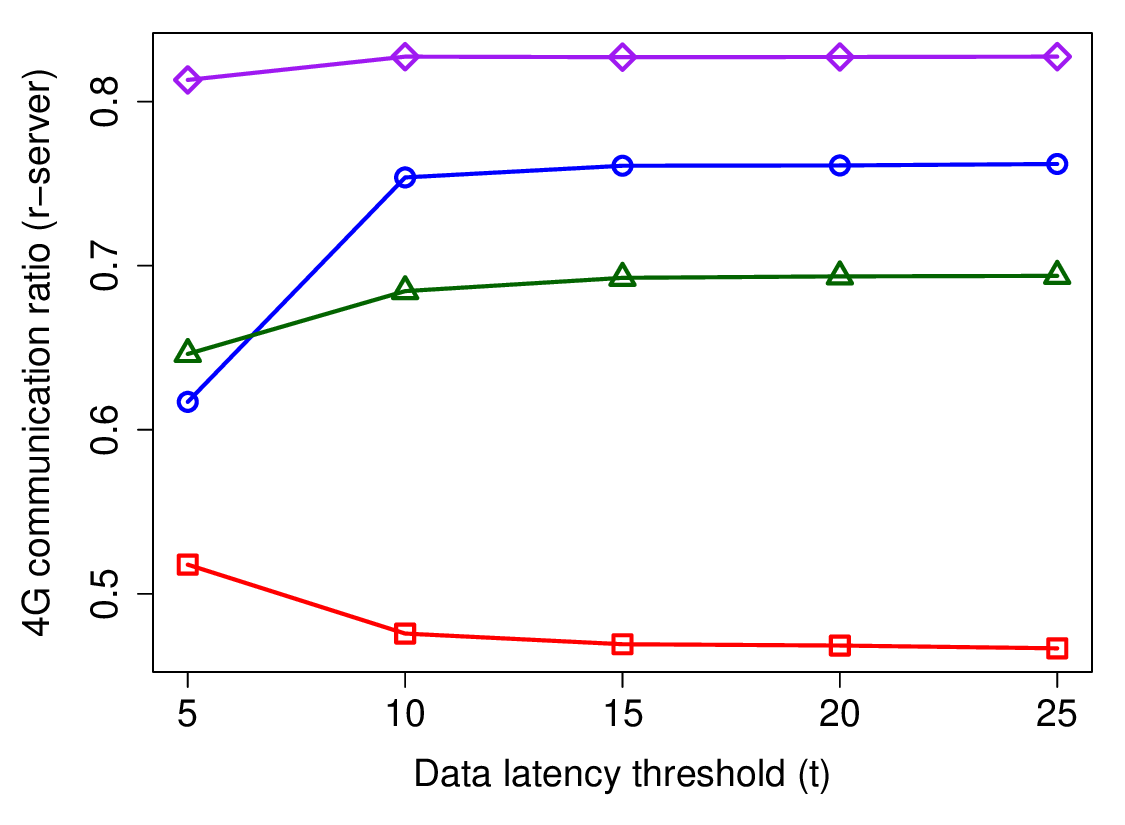} \label{fig:change_delay_1}} 
    \hfill
    \subfigure[Rate of $\delta$-delayed packets]{\includegraphics[width=0.64\textwidth]{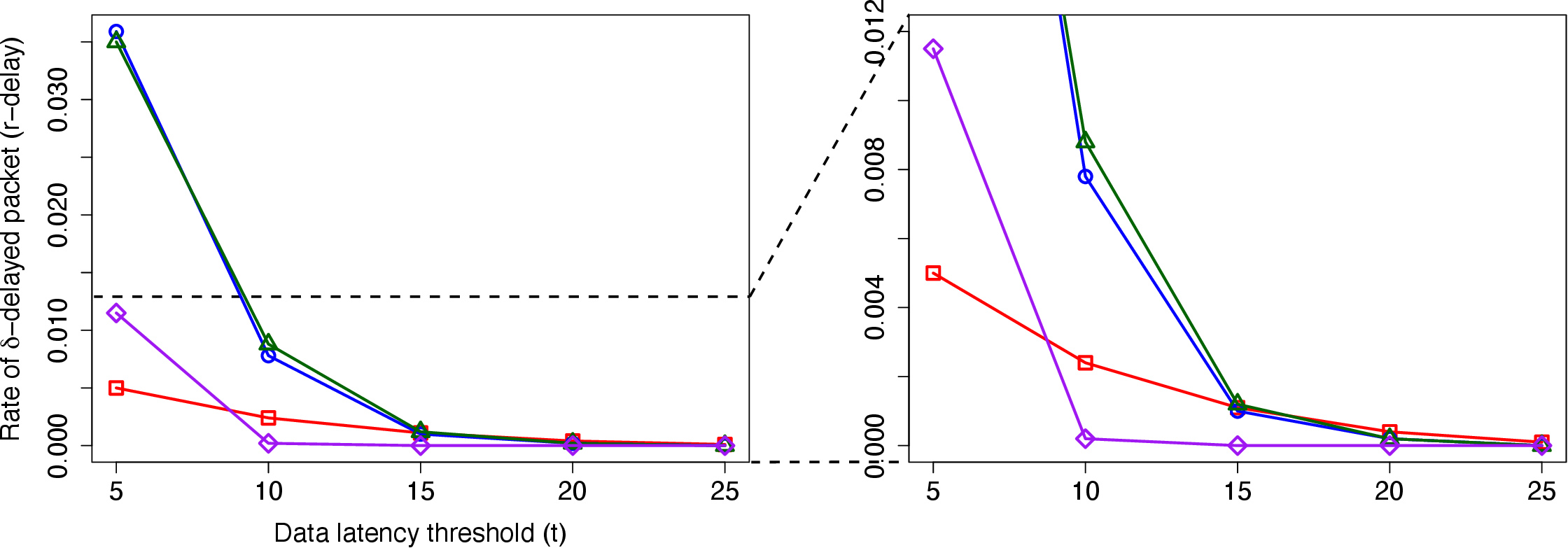} \label{fig:change_delay_2}} \\
    \subfigure[Rate of dropped packets]{\includegraphics[width=0.64\textwidth]{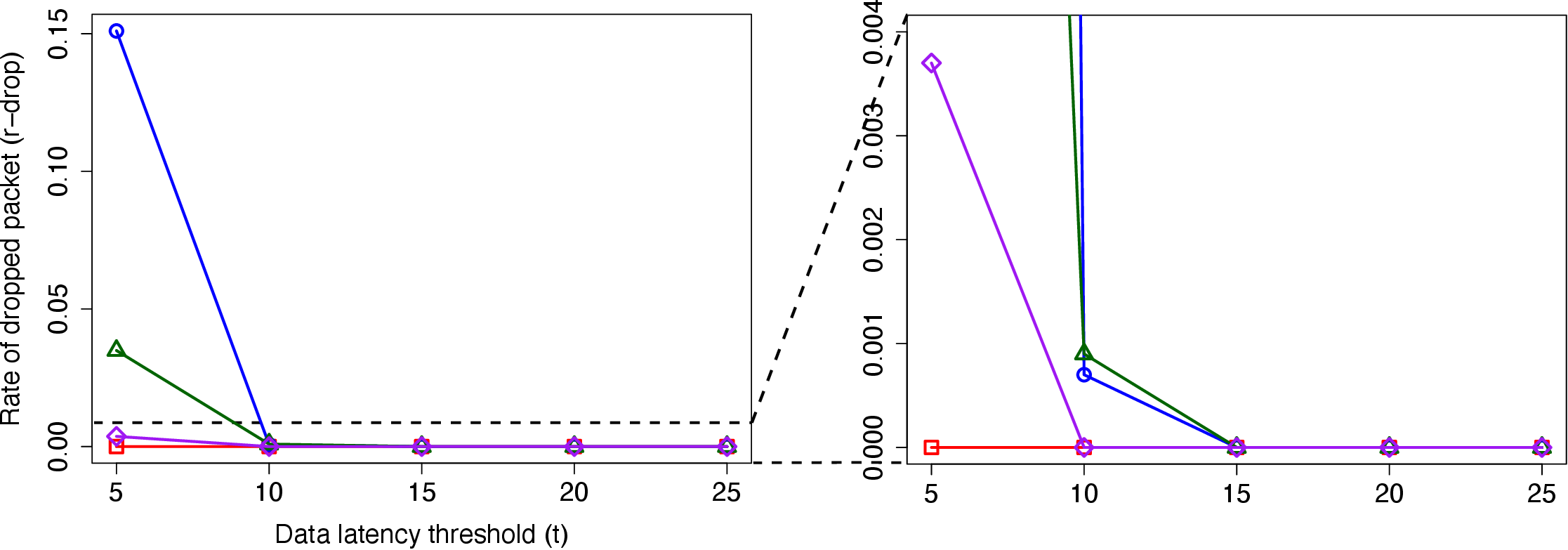} \label{fig:change_delay_3}}
    \caption{Impact of the data latency threshold ($\delta$). $\lambda_d$ is fixed to 1 time steps.}
    \label{fig:change_delay}
\end{figure*}

In addition, although the $r_{delay}$ of FP3 is lower than that of our proposed method in the case of $\delta=10$, it requires much more 4G cost than ours. Specifically, in all the experiments, our method requires the lowest communication cost ($r_{server}$), i.e., only around 50\% of packets travel through the 4G network. It is interesting to note that the \textit{offload-missed} issue mentioned in Section~\ref{sec:methodology} increases the communication cost, i.e., $r_{server}$ of the FP strategies is much higher than the expected value (which should be around $P_{server}$). 
The reason can be explained as follows. 
When a packet is decided to send to the nearest RSU/device but there is no available RSUs/devices around it, and the queue is full, the packet will be directly sent to server via the 4G communication channel. Thus, the 4G communication cost increases significantly. For example, in the FP1 and FP2, more than 60\% of packets use the 4G communication channel (although it is designed with a fixed possibility of 30\%). Those values are 80\% and 50\% for FP3, respectively. 

In summary, the results imply that the performance/cost of the baseline FB approach is sensitive to the environment/network due to its coarse fixed configuration. Thus, it requires effort to manually figure out the best configuration when implementing this method in the real world (for a given specific environment). 
By contrast, our proposed method learns the environment/network information to make the decision flexibly so that it can avoid both the packet-dropped issue and \textit{offload-missed} issue.

\subsection{Discussion}
\label{sec:discussion}
In the following, we investigate the impacts of the packet generation interval and the data latency threshold to our proposed method. 
\subsubsection{Impacts of the packet generation interval $\lambda_d$}
Figure~\ref{fig:generation_time} shows the impact of the packet generation interval, i.e., the frequency the devices measure the air quality indicators. In this evaluation, we set the data latency threshold to $5$ while changing the packet generation interval from $1 \to 5$ time steps. The result shows that there's a trivial impact of packet generation rate to our proposed method in both communication cost and performance. However, in the FB strategy, a higher number of messages are generated in a time unit, the higher possibility of a packet is dropped. On the other hand, when the packet generation interval is slow enough, instead of being dropped, a packet will be stored in a queue and, thus,  the rate of delayed packets is increased. In general, this trend also appears when the relative packet size over the capacity of the local queue is too big (that leads to a smaller number of the packet can be store in the queue until the queue is full).
Interestingly, the packet generation interval does not affect the communication cost of FB strategy.

\subsubsection{Impacts of the data latency threshold $\delta$}

    
Figure~\ref{fig:change_delay} illustrates the impact of the data latency threshold, e.g., the maximum latency required by application, to our proposed method. In this experiment, we fix the packet generation interval $\lambda_d = 1$ while changing the latency threshold $\delta$ from $5 \to 25$ times steps. As expected that the rate of $\delta$-delayed packets of both our proposed method and the baseline strategy decrease as the latency threshold increase because a packet has a longer time stay in the local edges, e.g., sensor or RSU. 
Furthermore, let us remind our strategy of the reward function in our Q-Learning method (as shown in the Formula (\ref{eq5})). When the data latency exceeds the threshold, there is a higher possibility of a packet to be directly sent to the server using the 4G communication channel.
As the results, when the latency threshold increases, the number of packets meets such condition and use the 4G communication channel becomes smaller.

%% file: 6-conclusion.tex


In this research, we focused on real-time mobile air quality monitoring systems which rely on devices mounted on vehicles. 
The devices continuously measure the air quality indicators and transfer them to the server via either 4G or Wi-Fi communication channels. 
We leveraged Q-learning to propose an opportunistic communication algorithm that minimizes the 4G communication cost while guaranteeing the data latency is under a predefined threshold.  
The experiment results showed that the proposed method can reduce 40-50\% of the 4G communication cost while ensuring the latency of 99.5\% packets smaller than the required threshold.


%% file: main.bbl
\begin{thebibliography}{10}

\bibitem{CHEN2019740}
Zhangjian Chen, Liangliang Cui, Xiaoxing Cui, Xinwei Li, Kunkun Yu, Kesan Yue,
  Zhixiang Dai, Jingwen Zhou, Guang Jia, and Ji~Zhang.
\newblock The association between high ambient air pollution exposure and
  respiratory health of young children: A cross sectional study in jinan,
  china.
\newblock {\em Science of The Total Environment}, 656:740--749, 2019.

\bibitem{LI20191304}
Xinwei Li, Xiao Zhang, Zhiqiang Zhang, Lianyu Han, Deping Gong, Jie Li, Ting
  Wang, Yanhua Wang, Sheng Gao, Huawei Duan, and Fanling Kong.
\newblock Air pollution exposure and immunological and systemic inflammatory
  alterations among schoolchildren in china.
\newblock {\em Science of The Total Environment}, 657:1304--1310, 2019.

\bibitem{PAM_Air}
{https://pamair.org/}.
\newblock Accessed: 2021-June.

\bibitem{KAIVONEN202023}
Sami Kaivonen and Edith C.-H. Ngai.
\newblock Real-time air pollution monitoring with sensors on city bus.
\newblock {\em Digital Communications and Networks}, 6(1):23 -- 30, 2020.

\bibitem{9322079}
Viet-Dung Nguyen, Phi~Le Nguyen, Trung~Hieu Nguyen, and Phan~Thuan Do.
\newblock A $\frac{1}{2}$ -approximation algorithm for target coverage problem
  in mobile air quality monitoring systems.
\newblock In {\em GLOBECOM 2020 - 2020 IEEE Global Communications Conference},
  pages 1--6, 2020.

\bibitem{DESOUZA2020102239}
Priyanka deSouza, Amin Anjomshoaa, Fabio Duarte, Ralph Kahn, Prashant Kumar,
  and Carlo Ratti.
\newblock Air quality monitoring using mobile low-cost sensors mounted on
  trash-trucks: Methods development and lessons learned.
\newblock {\em Sustainable Cities and Society}, 60:102239, 2020.

\bibitem{7553459}
K.~{Zhang}, Y.~{Mao}, S.~{Leng}, Q.~{Zhao}, L.~{Li}, X.~{Peng}, L.~{Pan},
  S.~{Maharjan}, and Y.~{Zhang}.
\newblock Energy-efficient offloading for mobile edge computing in 5g
  heterogeneous networks.
\newblock {\em IEEE Access}, 4:5896--5907, 2016.

\bibitem{platoon_1}
X.~Fan, T.~Cui, C.~Cao, Q.~Chen, and K.S. Kwak.
\newblock Minimum-cost offloading for collaborative task execution of
  mec-assisted platooning.
\newblock {\em Sensors}, 19(847), 2019.

\bibitem{platoon_3}
T.~Cui, Y.~Hu, B.~Shen, and Q.~Chen.
\newblock Task offloading based on lyapunov optimization for mec-assisted
  vehicular platooning networks.
\newblock {\em Sensors}, 19(4974), 2019.

\bibitem{platoon_2}
H.~{Wang}, X.~{Li}, H.~{Ji}, and H.~{Zhang}.
\newblock Federated offloading scheme to minimize latency in mec-enabled
  vehicular networks.
\newblock In {\em Proc. IEEE GLOBECOM Workshops}, pages 1--6, 2018.

\bibitem{8644315}
H.~{Wang}, X.~{Li}, H.~{Ji}, and H.~{Zhang}.
\newblock Federated offloading scheme to minimize latency in mec-enabled
  vehicular networks.
\newblock In {\em Proc. IEEE Globecom Workshops}, pages 1--6, 2018.

\bibitem{2_tier_1}
L.~{Feng}, W.~{Li}, Y.~{Lin}, L.~{Zhu}, S.~{Guo}, and Z.~{Zhen}.
\newblock Joint computation offloading and urllc resource allocation for
  collaborative mec assisted cellular-v2x networks.
\newblock {\em IEEE Access}, 8:24914--24926, 2020.

\bibitem{3-tier_1}
J.~{Zhao}, Q.~{Li}, Y.~{Gong}, and K.~{Zhang}.
\newblock Computation offloading and resource allocation for cloud assisted
  mobile edge computing in vehicular networks.
\newblock {\em IEEE Trans. Veh. Technol.}, 68(8):7944--7956, 2019.

\bibitem{8402110}
Y.~{Lin}, Y.~{Lai}, J.~{Huang}, and H.~{Chien}.
\newblock Three-tier capacity and traffic allocation for core, edges, and
  devices for mobile edge computing.
\newblock {\em IEEE Trans. Netw. Service Manag.}, 15(3):923--933, 2018.

\bibitem{watkins1992q}
Christopher~JCH Watkins and Peter Dayan.
\newblock Q-learning.
\newblock {\em Machine learning}, 8(3-4):279--292, 1992.

\bibitem{Khiem_Le_MEC}
Phi~Le Nguyen, Ren-Hung Hwang, Pham~Minh Khiem, Kien Nguyen, and Ying-Dar Lin.
\newblock Modeling and minimizing latency in three-tier v2x networks.
\newblock In {\em GLOBECOM 2020 - 2020 IEEE Global Communications Conference},
  pages 1--6, 2020.

\bibitem{jetcheva2003design}
Jorjeta~G Jetcheva, Yih-Chun Hu, Santashil PalChaudhuri, Amit Kumar, Saha
  David, and B~Johnson.
\newblock Design and evaluation of a metropolitan area multitier wireless ad
  hoc network architecture.
\newblock 2003.

\end{thebibliography}
